\documentclass[11pt]{article}

\usepackage{graphicx}

\usepackage{cite}

\begin{document}

%

% Title, Authors, Affiliations

% ===================

%

\title{\Large\bf
Higgs boson production in high energy proton-nucleus collisions}

\author{Andreas Sch\"{a}fer, Jian Zhou
 \\[0.3cm]
{\normalsize\it Institut f\"{u}r Theoretische Physik,Universit\"{a}t Regensburg, Regensburg, Germany}}

\maketitle

% headline

%\thispagestyle{fancy}

%\fancyhead[R]{\tt \jobname}

%

% Abstract

% ======

%

\begin{abstract}
\noindent
We study Higgs boson production from gluon-gluon fusion at mid-rapidity
in high energy proton-nucleus collisions.For this process the presently
still little known gluon distribution function $h_1^{\perp g}(x,k_{\perp})$ might give
a numerically relevant contribution.
We show by explicite calculation that
using CGC (color glass condensate) model input
the result obtained in
the naive $k_t$ factorization approach
matches the result obtained in the TMD factorization framework for a dilute medium.
We also verify the earlier finding~\cite{Sun:2011iw} that the $k_t$ factorization formalism for Higgs production
breaks down in a dense medium.
 In doing so we formulate a
hybrid model which allows one to treat such reactions theoretically.
\end{abstract}

%

% 1. Section: Introduction

% ==================

%

\section{Introduction}
\noindent In recent years the theoretical understanding of
transverse momentum dependent (TMD) parton distributions has made
tremendous progress, up to the point that their further
investigation can serve as one of the motivations for a new
electron-ion collider \cite{Boer:2011fh}. Still there are many
aspects which need further study. An especially interesting one is
relevant for Higgs-production at the LHC. In this case it was
realized that for $pp$ collisions in addition to the standard gluon
contribution one gets a contribution proportional to a little known
distribution function, usually refereed to as the distribution of
linearly polarized gluons~\cite{Mulders:2000sh,Anselmino:2005sh} ($h_{1}^{\perp g}$
in the notation of Ref.~\cite{Meissner:2007rx}). Therefore,
$h_{1}^{\perp g}$ has attracted a lot of attention recently. This
new distribution function is the only spin dependent gluon TMD for
an unpolarized nucleon/nucleus, and may be considered as the
counterpart of the quark Boer-Mulders function~$h_1^{\perp
q}(x,k_{\perp})$ \cite{Boer:1997nt}. However, in contrast to the
latter, $h_1^{\perp g}$ is time-reversal even implying that
initial/final state interactions are not needed for its
existence~\cite{Brodsky:2002cx,Collins:2002kn}. This distribution
function is of phenomenological interest, especially for small-$x$
physics at RHIC and LHC because a calculation in the saturation
model~\cite{Metz:2011wb} showed that its contributions are (at
small-$x$) as large as those proportional to the unpolarized gluon
distribution. Fortunately, it has been shown that $h_{1}^{\perp g}$
can be accessed, at least in principle, through measuring, e.g.,
azimuthal $\cos 2\phi$ asymmetries in processes such as jet or heavy
quark pair production in electron-nucleon scattering as well as
nucleon-nucleon scattering. Other promising observables are $\cos
2\phi$ asymmetries in photon pair production in hadron
collisions~\cite{Boer:2009nc,Boer:2010zf,Qiu:2011ai}. Such
measurements should be feasible at RHIC, the LHC, and a potential
future Electron Ion Collider
(EIC)~\cite{Anselmino:2011ay,Boer:2011fh} and could play an
important role to establish saturation effects. More recently, it
has been found that the linearly polarized gluon distribution may
affect the transverse momentum distribution of Higgs bosons produced
from gluon fusion for $p_{H\perp}\ll m_H$, where $p_{H\perp}$ and $
m_H$ are the Higgs transverse momentum and mass
respectively~\cite{Boer:2011kf,Sun:2011iw}. The authors of
Ref.~\cite{Boer:2011kf} proposed that the effect of linearly
polarized gluons on the Higgs transverse momentum distribution can
even be used, in principle, to determine the parity of the Higgs
boson experimentally.  Transverse momentum dependent factorization
 has been re-examined by taking into account the perturbative
gluon-radiation correction to $h_{1}^{\perp g}$~\cite{Sun:2011iw}.
The complete TMD factorization results for Higgs boson production
are consistent with earlier findings based on the
Collins-Soper-Sterman (CSS) formalism~\cite{Catani:2010pd} and
soft-collinear-effective theory~\cite{Mantry:2009qz}. Also, the
transverse momentum resummation formalism applied to di-photon
production in $pp$ collisions~\cite{Nadolsky:2007ba} is closely
related.

Besides their obvious phenomenological interest, these investigations of Higgs production are
also interesting from a more theoretical point of view. The theoretical description of
transverse momentum dependent processes unavoidably includes gauge links in one or the other form.
While the starting expressions for the different approaches look often quite different
with respect to these gauge links, it turned out that quite often the resulting cross sections can
be mapped onto one another in some approximation and for a suitable kinematic window.
In the following we will discuss three such approaches, TMD factorization, $k_t$ factorization
and a new hybrid approach.\\

{\bf TMD factorization}\\
We will use the term TMD factorization in the sense of
\cite{Collins:2011zzd}, which defines hard and soft factors, such that,
e.g., the Higgs boson production cross section with $P_{\perp}\ll M$ reads
\cite{Sun:2011iw}
\begin{eqnarray}
\frac{d^3\sigma(M^2,P_{\perp},y)}{d^2P_{\perp}dy} &=& \sigma_0 \int
d^2\vec k_{1\perp} d^2\vec k_{2\perp}d^2\vec\ell_{\perp}~
\delta^{(2)} (\vec k_{1\perp}+\vec k_{2\perp}+\vec \ell_{\perp}-\vec
P_{\perp})
\nonumber \\
&\times
&\Bigl\{x_1g(x_1,k_{1,\perp})~x_2g(x_2,k_{2,\perp})~S(\ell_{\perp},\mu\rho)~
H(M^2,\mu\rho)
\nonumber \\
&+& \left( \frac{2(k_{1\perp}\cdot
k_{2\perp})^2}{k_{1\perp}^2k_{2\perp}^2}-1\right) ~x_1h_1^{\perp
g}(x_1,k_{1,\perp})~x_2h_1^{\perp g}(x_2,k_{2,\perp})
\nonumber \\
&\times& S_h(\ell_{\perp},\mu\rho)~
H_h(M^2,\mu\rho) \Bigr\}
\end{eqnarray}
with the soft factors $S$ and $S_h$ and the hard interaction factors
$H$ and $H_h$. TMD factorization is probably the formally most
complete and reliable scheme, but often also the calculational most
demanding. For specific questions other schemes might be more
economic. For example for all-order proofs of factorization SCET is
a promising alternative \cite{Manohar:2012jr} and for qualitative
phenomenological analysis TMD factorization promisees a substantial
simplification. Within TMD factorization there also exist different
approaches. To be specific we use this term for the formulation of
Collins et al. for which it is crucial to define the gauge links
slightly off the light-cone. Within SCET it is possible to keep the
gauge links on the light-cone \cite{GarciaEchevarria:2011rb}. In
principle both approaches should give consistent results for
physical observables when expanded in an appropriate manner, but it
is
non-trivial to map, e.g., evolution in both schemes onto one another.\\

{\bf $k_t$ factorization}\\
In contrast, the naive $k_t$ factorization scheme invokes some approximations.
In this formulation there is no linearly polarized gluon distribution function which
is equivalent to the statement that it has to have the same functional form as the normal
unpolarized gluon distribution such that it cannot be discerned.
This fact demonstrates clearly that $k_t$ factorization can in general not be a good
approximation because within CGC framework both gluon distributions could
differ substantially for $Q_s \gg  k_{\perp} $ as was first
noticed in~\cite{Metz:2011wb}. This can be best seen from the following expressions for
the Weizs\"{a}cker-Williams (WW) unpolarized gluon distribution denoted by $G_{WW}(x,k_{\perp})$ and the WW type
linearly polarized gluon distribution $h^{\perp g}_{1,WW}(x,k_{\perp})$ derived in the CGC formalism
~\cite{Kovchegov:1996ty,JalilianMarian:1996xn,Metz:2011wb},
\begin{eqnarray}
x G_{WW}(x,k_\perp) \!\!\!
& = & \!\!\! \frac{N_c^2-1 }{N_c} \frac{S_\perp }{4 \pi^4 \alpha_s} \int d^2 r_\perp \,
e^{-i \vec k_\perp \cdot \vec r_\perp} \,
\frac{1}{r_\perp^2} \bigg( 1 - e^{ - \frac{r_\perp^2 Q_s^2}{4}} \bigg) \,,
\\
x h^{\perp g}_{1,WW}(x,k_\perp) \!\!\! &=& \!\!\!
\frac{N_c^2-1 }{8 \pi^3} \, S_\perp \int d r_\perp \,
\frac{J_2 (k_\perp r_\perp)}{\frac{1}{4 \mu_A} r_{\perp} Q_s^2}
\bigg( 1 - e^{ - \frac{r_\perp^2 Q_s^2}{4}} \bigg) \,.
\label{eq:1}
\end{eqnarray}
Here $S_\perp$ is the transverse area of the target nucleus, and $k_{\perp} \equiv |\vec{k}_{\perp}|$.
$Q_s^2 = \alpha_s N_c \mu_A{\rm ln} \left [ 1 / (r_\perp^2 \Lambda_{QCD}^2) \right ]$ is the
gluon saturation scale with $\mu_A$ being a common CGC parameter.
Note that our convention for $h^{\perp g}_{1,WW}$  differs from that in
Ref.~\cite{Metz:2011wb} by a factor 1/2. In general, both gluon distributions are different though they
 become identical in the dilute region, i.e. for $k_{\perp}\gg Q_s$\\

{\bf hybrid approach}\\
Our main strategy is to perform the calculations for proton-nucleus reactions in an
approach where the nucleus is treated in the CGC framework~
\cite{McLerran:1993ni,Mueller:2001fv,Iancu:2002xk,Gelis:2010nm}, which
effectively takes soft gluons into account, and
the proton in the so-called Lipatov approximation
~\cite{Kuraev:1977fs,Gribov:1984tu,Catani:1990eg,Collins:1991ty}.
We restrict ourself to studying Higgs production in the plain
MV model~\cite{McLerran:1993ni} without considering
small $x$ evolution effects although the latter could be done in
principle~\cite{Dumitru:2010mv,Dominguez:2011gc,Dominguez:2011br,Dumitru:2011vk,Iancu:2011ns,Jmarian,Iancu:2011nj}
by solving the general JIMWLK evolution equation~\cite{JalilianMarian:1997gr} for quadrupole operator.
Neglecting evolution should be a good approximation because the MV model is valid in the range
$x \approx 0.01-0.1$, which is the relevant kinematical regime for Higgs boson production at LHC,
while radiative corrections only become important below a certain scale $x \approx 0.01$.\\

This article is organized as follows.
In Sec. II, we introduce the hybrid approach and use it to reproduce the well known
result for soft gluon production in $pA$ collisions.
The Higgs boson production in $pA$ collisions is computed in the same approach in Sec III.
We also demonstrate consistency between the results obtained in
TMD factorization and $k_t$ factorization in the dilute region within the CGC model.
We summary our paper in Sec. IV.

\section{Soft gluon production in proton-nucleus collisions}

\noindent
In this section, we introduce a hybrid approach and reproduce the well known result for
gluon production in high energy proton-nucleus scattering
~\cite{Kopeliovich:1998nw,Kovchegov:1998bi,Kharzeev:2003wz,Dumitru:2001ux,Blaizot:2004wu,Gelis:2005pt}.
In the next section we shall generalize this approach to Higgs production.
Let us first consider the general case of soft gluon production
\begin{eqnarray}
A(P_A)+p(P_B) \rightarrow g(l)+X \ .
\end{eqnarray}
We assume that the nucleus is moving with a velocity very close to the speed of light into the positive
$z$ direction,
while the proton is moving in the opposite direction.
It is convenient to use light-cone coordinates for which $P_A^\mu=P_A^+p^\mu$ and $P_B^\mu=P_B^- n^\mu$
with $p = (1, 0, 0, 0)$ and $n = (0, 1, 0, 0)$.  The corresponding partonic subprocess is represented by
$g_A(k_1)+g_p(k_2) \rightarrow g(l) $, where $k_1^\mu=x_1P_A^+ p^\mu+ k_{1\perp}^\mu$ denotes the total momentum
carried by multiple gluons from  the nucleus, and
$k_2^\mu=x_2 P_B^-n^\mu+k_{2\perp}^\mu$ is the momentum of the gluon from the proton.
We chose to work in the light-cone gauge of the proton $(A^-=0)$, for which the polarization tensor of
a produced gluon
is  given by,
\begin{eqnarray}
\sum \epsilon^\mu \epsilon^{*\nu }= -g^{\mu \nu}+ \frac{p^\mu l^\nu+p^\nu l^\mu}{p \cdot l} \ .
\end{eqnarray}
As mentioned above, to facilitate our calculation, a hybrid strategy has been adopted, in
which the nucleus is treated in the CGC model,
while on the side of the dilute projectile proton one makes the so-called Lipatov
approximation~\cite{Kuraev:1977fs,Gribov:1984tu,Catani:1990eg,Collins:1991ty}.
At small $x$ the gluon radiation cascade shows a strong ordering in rapidity.
Or in other words, color source carries much larger rapidity than that radiated gluon does.
It has been shown that a fast moving color source can be treated as eikonal line in the strongly rapidity ordered
region. The operation of introducing these eikonal lines is refereed to as Lipatov approximation~\cite{Collins:1991ty}.
Its validity has been confirmed also by solving the classical Yang-Mills
equation~\cite{Dumitru:2001ux,Blaizot:2004wu,Gelis:2005pt}.
In these calculations, the gluon field induced inside a proton by a weak color source and weak color source itself are treated
as a small parameter when solving classical Yang-Mills equation  perturbatively.
An analytic solution for the gauge field was obtained in lowest order of
the incoming gluon field in various gauges~\cite{Dumitru:2001ux,Blaizot:2004wu,Gelis:2005pt}.

For the process of gluon production in $pA$ collisions, the relevant eikonal line is
the past-pointing one which is built up through initial state interactions between the
color source inside the proton and the background gluon field.
The interaction between the classical gluon field and the final state gluon emitted from
the color source inside the proton does not change this general statement because
the imaginary part of the scattering amplitude cancels between the different cut diagrams
once the final states are integrated out.
The prescription to treat the eikonal propagator is fixed by this choice.
The relevant Feynman rules, illustrated in Fig.\ 1, were given in Ref.~\cite{Collins:1991ty}.
Note that the prescription for past-pointing eikonal propagators differs from that for
future-pointing eikonal lines.

\begin{figure}[t]
\begin{center}
\includegraphics[width=12cm]{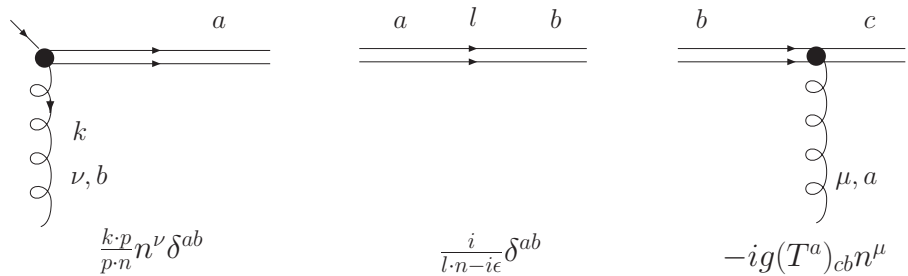}
\end{center}
\vskip -0.4cm \caption{ Feynman rules for the eikonal line, which is  represented by a double line. $a,b,c$ denote color indices.}
\label{efig1}
\end{figure}

It is worthwhile to mention that to preserve gauge invariance one has to take both, gluon fusion and
the interaction between the color source inside the proton and the strong classical gluon field of
the large nucleus into account.
This is due to the fact that the incoming gluon from the proton is off-shell and off-shell quantities are, in general, not gauge invariant.
Both types of interaction are shown in Fig.\ 2.

The multiple scattering between incoming gluon (or eikonal line) and the classical color
field of the nucleus can be readily resumed to all orders as has been done for the scattering of a
quark by a background gluon field~\cite{Balitsky:1995ub,McLerran:1998nk}.
The resumed multiple scattering gives rise to a path-ordered gauge factor along the straight line
that extends in $x^-$ from minus infinity to plus infinity. More precisely, for a gluon or eikonal line  with incoming momentum
being $k$ and outgoing momentum being $k+q$,
the path-ordered gauge factor reads,
\begin{equation}
 2 \pi  \delta(q^-)
 p^\mu [U-1](q_\perp) \,,
\end{equation}
with
\begin{equation}
[U-1](q_\perp)
=\int d^2 x_\perp e^{-  i \vec q_\perp \cdot \vec x_\perp} [U(x_\perp)-1] \,,
\end{equation}
and
\begin{equation}
U(x_\perp)= \langle {\cal P} e^{-ig \int_{-\infty}^{+\infty} dx^- A^+(x^-, \ x_\perp)} \rangle_A \, ,
\end{equation}
where $A^+=A^+_c t^c$ is the gluon potential in the adjoint representation and
$t^c_{ab}=-if_{abc}$ .\\

\begin{figure}[t]
\begin{center}
\includegraphics[width=12cm]{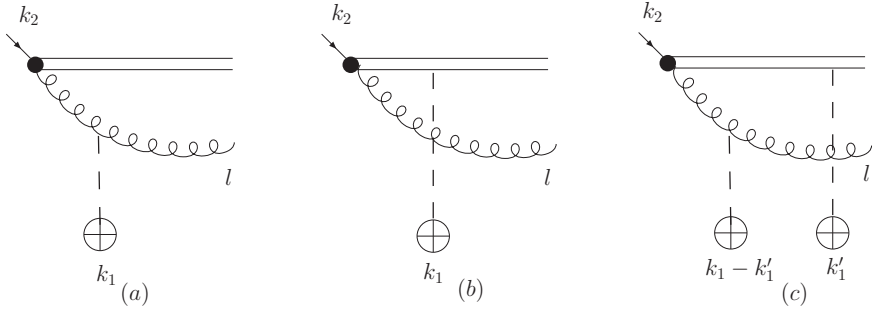}
\end{center}
\vskip -0.4cm \caption{ The diagrams contributing to gluon production. The dash lines represent
the re-summed interactions of the incoming gluon or color source inside the proton
with the classical color field of the nucleus.}
\label{efig2}
\end{figure}
We use this as a building block to compute the amplitude for gluon production in high energy $pA$ collisions.
It is easy to verify that the contribution from the third diagram vanishes because both  ${k_1'}^+$ poles
are located in the same half plane.
Consequently, we are left with the contributions from the first two diagrams.
The calculation of these two diagrams is straightforward.
Collecting all pieces, the differential cross section reads,
\begin{eqnarray}
\frac{d \sigma}{d^2 l dy} &=& \frac{\pi}{(N_c^2-1)  l_\perp^2 } \int
\frac{2 d^2 k_{1\perp}}{(2\pi)^3} k_{1\perp}^2 x_2 g (x_2,k_{2\perp} )
[ U-1  ] (k_{1\perp}) [ U^\dag-1  ] (k_{1\perp})
\nonumber \\
&=&
\frac{1}{(N_c^2-1)  l_\perp^2 }
\int \frac{d^2 k_{1\perp}}{(2\pi)^2} k_{1\perp}^2 x_2 g (x_2,k_{2\perp} )
U (k_{1\perp})  U^\dag (k_{1\perp}) \ ,
\end{eqnarray}
where $g(x_2,   k_{2\perp} \equiv | \vec k_{2\perp} |= |\vec l_\perp-\vec k_{1\perp}|)$  denotes the un-integrated gluon
distribution of a proton,
and $y$ is the rapidity of the produced gluon. The factor $2 /(2\pi)^3 $ associated with phase space
integration is chosen such that for single gluon target,
$\int \frac{2 d^2 k_{1\perp}}{(2\pi)^3} \frac{k_{1\perp}^2}{g^2 N_c}\langle U (k_{1\perp})  U^\dag (k_{1\perp}) \rangle_{\rm gluon}=x_1 \delta(1-x_1)$
at lowest non-trivial order (see, for example, Ref.~\cite{Iancu:2002xk}).
To obtain the above result, we  have defined the normalization factor and
the flux factor to be
$ k_{2\perp}^2/(2 k_2 \cdot p (N_c^2-1))$ and $1/(2 k_2 \cdot p)$, respectively,
rather than $k_{1\perp}^2 k_{2\perp}^2/(4 x_1 x_2 P_A \cdot P_B  (N_c^2-1)^2)$, $1/(4x_1 x_2 P_A \cdot P_B)$
used in Ref.~\cite{Collins:1991ty}, since the Lipatov approximation is only justified for
the proton side.

The next step is to compute the expectation value of a Wilson line in the
plain McLerran-Venugopalan model~\cite{McLerran:1993ni}.
By averaging over color sources with a Gaussian distribution,
one finds~\cite{Gelis:2001da,Blaizot:2004wv,Fukushima:2007dy},
\begin{equation}
\left \langle {\rm Tr} U(R_\perp+r_\perp)U^\dag(R_\perp) \right \rangle_A
=(N_c^2-1) \exp \left \{ \frac{-Q_s^2 r_\perp^2}{4} \right \}  \ .
\end{equation}
To proceed further, one notices that the dipole type gluon distribution in the adjoint representation
is given by~\cite{Kharzeev:2003wz,Braun:2000wr,Braun:2000bh},
\begin{equation}
 x G_{DP}(x, k_\perp)=
\frac{C_F S_\perp}{ 2 \pi^2 \alpha_s}k_\perp^2
\int \frac{d^2 r_\perp}{(2\pi)^2} \, e^{-i \vec k_\perp \cdot \vec r_\perp} \,
e^{-\frac{Q_{s}^2 r_\perp^2}{4}} \ .
\end{equation}
With the help of the above two equations, the differential cross section can be written as,
\begin{equation}
\frac{d \sigma}{d^2 l dy}
=\frac{4 \pi^2 \alpha_s N_c}{(N_c^2-1)  l_\perp^2}
\int d^2 k_{1\perp} x_2 g (x_2,k_{2\perp})
 x_1 G_{DP} \left (x_1,k_{1\perp} \right ) \,
 \label{1}
\end{equation}
which is in full agreement with the results of earlier
calculations
~\cite{Kopeliovich:1998nw,Kovchegov:1998bi,Kharzeev:2003wz,Dumitru:2001ux,Blaizot:2004wu,Gelis:2005pt,Avsar:2012hj}.
To see this, one has to identify $\pi x_2 g (x_2,k_{2\perp}) \equiv  \partial[x_2 g (x_2,Q^2)]/\partial Q^2|_{Q^2=k_{2\perp}^2}$
and $\pi x_1 G_{DP}  (x_1,k_{1\perp}) \equiv  \partial[x_1 G(x_1,Q^2)]/\partial Q^2|_{Q^2=k_{1\perp}^2}$, where
$g (x_2,Q^2)$ and $G(x_1,Q^2)$ are the integrated gluon distributions of proton and nucleus.
In particular, Eq.[12] demonstrates that the gluon production
cross section can be expressed in terms of the gluon distribution in a rather straightforward manner.
In the next section, we apply this hybrid approach to calculate the Higgs boson production in $pA$
collisions.
In contrast to gluon production for which the dipole gluon distribution appears in the cross section,
one has to use Weizs\"{a}cker-Williams gluon distributions in the calculation for Higgs boson production,
as we show below.

\section{Higgs boson production in proton-nucleus collisions}
Now we turn to Higgs boson production in proton-nucleus collisions.
Let us start by introducing the matrix element definition for gluon TMDs
that generate the Higgs transverse momentum distribution
~\cite{OITS-166,hep-ph/0503015,Mulders:2000sh,Meissner:2007rx},
\begin{eqnarray} \label{e:ww}
&& \!\!\! \int \frac{d r^- d^2 r_\perp}{(2\pi)^3 P^+} \,
e^{-ix_1P^+r^- + i \vec k_{1\perp} \cdot \vec r_\perp}
\langle A | F^{+i}(r^- + y^-, r_{\perp} + y_{\perp}) \, L_{r+y}^\dag  \, L_{y} \,
F^{+j}(y^-,y_\perp) |A \rangle
\nonumber\\ && \!\!\!\!\!\!\!\!\! =
\frac{\delta_{\perp}^{ij}}{2} \, x_1 G_{WW}(x_1, k_{1\perp}) +
\bigg(\hat k_{1\perp}^i \hat k_{1\perp}^j - \frac{1}{2} \delta_{\perp}^{ij} \bigg)
x_1 h^{\perp g}_{1,WW}(x_1, k_{1\perp}) \,,
\end{eqnarray}
where $\hat{k}_{1\perp}^i = \vec k_{1\perp}^i/ k_{1\perp}$, and $\delta_{\perp}^{ij}=-g^{ij}+(p^in^j+p^jn^i)/p \cdot n$.
The gauge link extends to the past: $L_{y} = \mathcal{P} \, e^{-ig \int_{\infty^-}^{y^-} d \zeta^- A^+(\zeta^-, y_{\perp})} $.
The two leading power gluon TMDs $G_{WW}(x_1, k_{1\perp})$, $h^{\perp g}_{1,WW}(x_1, k_{1\perp})$ are the
usual WW type unpolarized TMD gluon distribution and WW type TMD distribution of linearly polarized gluons
respectively.
As we show below, both gluon TMDs contribute to the differential cross section for Higgs
production.

 Higgs boson production through gluon-gluon fusion has been studied in the context of
the $k_t$ factorization formalism~\cite{Lipatov:2005at,Hautmann:2002tu,Pasechnik:2006du}.
However, the authors of Ref.~\cite{Sun:2011iw} argued that $k_t$ factorization can break down in a dense medium
where the WW type linearly polarized gluon distribution is different from the usual gluon distribution.
Then the Higgs production cross section cannot be expressed only in terms of the usual gluon distribution
and TMD and $k_{\perp}$ factorization give different results.
However, they also argue that one should be able to modify $k_t$ factorization to establish an effective
TMD factorization at small x. One possibility to do so was recently proposed in
Refs.~\cite{Dominguez:2010xd,Dominguez:2011wm}.
Generally speaking, at very small $x$ higher twist contributions are as important as the leading twist
ones because of the high gluon density.
Therefore, in order to arrive at the mentioned effective TMD factorization, an analysis including
all higher twist contributions is crucial.
For the unpolarized case it was shown that the results for two-particle
correlations in high energy scattering using the proposed effective TMD factorization
are in agreement with the results obtained by extrapolating the CGC calculation to the correlation
limit~\cite{Dominguez:2010xd,Dominguez:2011wm}, where the transverse momentum imbalance between
the two final state particles (or jets) is much smaller than the individual transverse momenta.
By applying a corresponding power counting in momentum space in the correlation limit, a complete matching
between the effective TMD factorization and the CGC formalism has
also been found in the polarized case~\cite{Metz:2011wb}.
Later, a calculation performed in position space led to the same results~\cite{Dominguez:2011wm}.

Inspired by Ref.~\cite{Sun:2011iw}, we carry out an explicit calculation for Higgs boson production
in $pA$ collisions using the CGC formalism and verify the conjecture that the effective TMD factorization
and CGC approach provide the same result in the dense medium region, while
the $k_t$ factorization is only valid in the dilute region.
The starting point is  the effective Lagrangian for Higgs boson production,
\begin{eqnarray}
{\cal L }_{eff}=-\frac{1}{4}g_\phi \Phi F_{\mu \nu}^a F^{a \mu\nu}  \ ,
\end{eqnarray}
which is valid in the heavy top quark limit, where $\Phi$ is the scalar field and $F_a^{\mu\nu}$ the gluon field strength.
$g_\phi$ is the effective coupling. The same effective
Lagrangian has also been used  to study gluon saturation in semi-inclusive DIS off
large nuclei~\cite{Kovchegov:1998bi}.
From the above Lagrangian, we can read off the basic vertices for the Higgs boson coupling
to gluon.
The corresponding Feynman rule for a Higgs boson coupling to two off-shell gluons carrying the
momenta $k_1$, $k_2$, and color indices $a$, $b$ is given by
\begin{eqnarray}
i g_\phi \delta_{ab} \left ( g^{\mu \nu} k_1 \cdot k_2  -k_1^\mu k^\nu_2 \right )  \ .
\end{eqnarray}
We argue that  Higgs boson production through a multiple gluons fusion process is
not enhanced by saturation effects for the following reasons:
First, the dominant component of the classical gauge field $A^+$ is generated by
the color sources inside the nucleus in a reasonably local way~\cite{Kovchegov:1998bi}.
Second, the coherence length for Higgs boson production is very small due to the large top quark mass.
Therefore, to calculate the transverse momentum dependent cross section for Higgs boson production
it is sufficient (in the saturation region) to use the effective vertex given above.

\begin{figure}[t]
\begin{center}
\includegraphics[width=15cm]{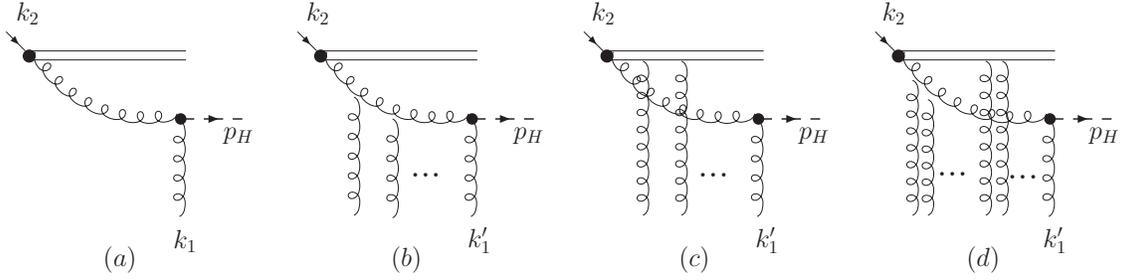}
\end{center}
\vskip -0.4cm \caption{ The generic diagrams contributing to Higgs boson production.
The multiple scattering of the incoming gluon or color source inside the proton
by the classical gauge field have to be re-summed to all orders.
$p_H$ denotes the Higgs boson momentum. }
\label{efig3}
\end{figure}

The relevant diagrams for Higgs boson production are shown in Fig.\ 3.
Before the incoming gluon from the proton combines with a gluon from the classical gauge field
of the nucleus to form a Higgs boson, this gluon has
initial state interactions with background gluons as it passes through the nucleus.
Although they do not modify the integrated production rate, the initial state interactions change
the transverse momentum distribution of the produced Higgs boson. The initial state interaction
between the color source inside the proton
and the classical gluon field of the nucleus should in principle also be taken into account.
However, such initial state interaction, shown in Fig.\ 3(c) and Fig.\ 3(d)
vanish because the  ${k_1'}^+$ poles are in the same half plane. Only Fig.\ 3(a) and Fig.\ 3(b) give
non-vanishing contributions.
Resuming gluon re-scattering to all orders, as illustrated in Fig.\ 3(b), and combining it with
the contribution from Fig.\ 3(a), one obtains the production amplitude,
\begin{eqnarray}
{\cal M}= g_\phi (k_2 \cdot p) \int \frac{d^2 k_{1\perp}'}{(2 \pi)^2} \left ( \frac{\vec k_{2\perp} \cdot \vec k_{1\perp}'}{k_{2\perp}^2} \right  )
 \langle   L(k_{1\perp}-k_{1\perp}')  A^+(k_{1\perp}')  \rangle_A   \phi_p(x_2,k_{2\perp})
\end{eqnarray}
where the $k_1^{'+}$ and $k_1^{'-}$ components have been integrated out.
$k_2=p_{H}-k_1$ denotes the momentum of the gluon from the proton with
$p_{H}$ being the Higgs momentum.
$\phi_p(x_2,k_{2\perp})$ represents the probability amplitude  for finding a gluon carrying a
certain momentum inside the proton, with
$x_2g(x_2,k_{2\perp})=\phi_p\phi_p^*$. It is convenient to introduce the gauge potential in
impact parameter space,
\begin{eqnarray}
A^+(k_{1\perp}')=\int dy^- d^2 y_\perp e^{ix_1P^+ y^- -i   \vec k_{1\perp}' \cdot \vec y_\perp } A^+(y^-,y_\perp) \ .
\end{eqnarray}
Replacing the exponential $e^{ix_1'P^+ y^-}$ by $e^{ix_1P^+ y^-}$ in the above formula is justified in
the leading logarithm approximation.
Moreover, one notices that $ L(k_{1\perp}-k_{1\perp}')$ is a Wilson line which starts from
$\xi^-$ being  minus infinity and ends at the space-time point $y^-$,
\begin{equation}
 L(k_{1\perp}-k_{1\perp}')
=\int d^2 \xi_\perp e^{-i (\vec k_{1\perp}-\vec k_{1\perp}') \cdot \vec \xi_\perp}
 {\cal P} e^{-ig \int_{-\infty^-}^{y^-} d\xi^- A^+(\xi^-, \ \xi_\perp)} \ .
\end{equation}
We proceed by partial integration and by performing the integral over $k_{1\perp}'$.
The amplitude then can be written as,
\begin{eqnarray}
{\cal M} &=& g_\phi (k_2 \cdot p ) \frac{-i \vec k_{2\perp , i}}{k_{2\perp}^2} \int d^2 y_\perp dy^- e^{ix_1P^+y^--i \vec k_{1\perp} \cdot \vec y_\perp}
\nonumber\\ && \times
 \langle  {\cal P} e^{-ig \int_{-\infty^-}^{y^-} d\xi^- A^+(\xi^-,y_\perp)}   F^{+i}(y^-,y_\perp)  \rangle_A   \phi_p(x_2,k_{2\perp}) \ .
\end{eqnarray}
Using the same normalization and flux factors as in the previous section, the differential cross section becomes,
\begin{eqnarray}
\frac{d^3 \sigma}{d^2 p_{H\perp} dy} \!\!\!  & = & \!\!\!
 \frac{\pi g_\phi^2}{32} \int \frac{2 d^2 k_{1\perp} }{(2\pi)^3 } \frac{\vec k_{2\perp,i} \vec k_{2\perp,j}}{k_{2\perp}^2}
x_2g(x_2,k_{2\perp}) \int d^3 r d^3y
\nonumber\\ && \ \ \ \ \
 \times \
 e^{-ix_1P^+r^- + i \vec k_{1\perp} \cdot \vec  r_\perp}
\langle  F^{+i}(r^- + y^-, r_{\perp} + y_{\perp}) \, L_{r+y}^\dag  \, L_{y} \,
F^{+j}(y^-,y_\perp)  \rangle_A
\nonumber\\ &=& \!\!\!
 \sigma_0 \int d^2 k_{1\perp}
x_2g(x_2,k_{2\perp}) x_1 \left [
  G_{WW}(x_1, k_{1\perp})+\left (2 (\hat k_{1\perp} \cdot  \hat k_{2\perp})^2 -1 \right )
 h^{\perp g}_{1,WW}(x_1, k_{1\perp}) \right ]
\end{eqnarray}
where $ \sigma_0=\frac{\pi g_\phi^2}{64} $ is the leading-order cross section for scalar-particle production
from two gluons.
Here, $y$ and $P_{H\perp}$ are rapidity and transverse momentum of the Higgs particle.
In the second step of above derivation,
we have made use of the normalization conditions for the average over the CGC wave
function: $\langle 1 \rangle_A=1$; and for the
nuclear state $|P \rangle$ carrying momentum $P$:
$ \langle P' |P \rangle=2 P^+(2\pi)^3 \delta(P^+-{P'}^+)\delta^2(P_\perp-P'_\perp)$~\cite{Dominguez:2011wm}.

As observed in Ref.~\cite{Sun:2011iw}, one automatically takes into account the contribution from the linearly polarized gluon TMD
in $k_t$ factorization. In other words, the usual unpolarized gluon distribution of the proton is the same
as its linearly polarized gluon distribution in the Lipatov approximation.
By noticing this fact, one finds that
the differential cross section computed in TMD factorization~\cite{Boer:2011kf,Sun:2011iw}
completely agrees with that derived in the CGC approach.
It is also worthwhile to mention that the gluon distributions entering the cross section for Higgs boson
production are the WW type distributions as expected. Furthermore, in order to compare this with results from the
$k_t$ factorization formalism, we recall that the WW type distribution $G_{WW}$ and
$ h^{\perp g}_{1,WW}$  become identical in the dilute region where $k_{1\perp} \gg Q_s$, but differ in a dense medium~\cite{Metz:2011wb}.
Therefore,  in the dilute region, the differential cross section can be simplified to
\begin{eqnarray}
\frac{d^3 \sigma}{d^2 p_{H\perp} dy}   =\sigma_0 \int d^2 k_{1\perp}
x_2g(x_2,k_{2\perp}) x_1 G_{WW}(x_1, k_{1\perp})2 (\hat k_{1\perp} \cdot  \hat k_{2\perp})^2
\end{eqnarray}
which agrees  with the well known result obtained from the $k_t$ factorization
approach~\cite{Lipatov:2005at,Hautmann:2002tu} at low Higgs transverse momentum.
Thus we conclude that also for Higgs production in $pA$
the $k_t$ factorization formula is only valid in the dilute region.
Similar conclusions have also been drawn for $\eta'$ meson production~\cite{FillionGourdeau:2008ij}
and heavy quark pair production~\cite{Blaizot:2004wv} in $pA$ collisions.

\section{Summary}

We developed a hybrid approach for calculating  particles production at
central rapidity in $pA$ collisions, in which
the dense target nucleus is treated in the color glass condensate model,
while on the side of the dilute projectile proton
the Lipatov approximation was used.
As a test of the method, we first reproduced the well known result for soft gluon
production in $pA$ collisions using this hybrid approach.
Then, we derived the differential cross section for Higgs boson production from gluon
fusion in $pA$ collisions.
It turned out that the result obtained in our hybrid approach is completely equivalent to that
computed in TMD factorization.
In the low-density limit of $pA$ collisions, we also recover the result of the naive $k_t$ factorization
formalism that describes Higgs boson production in $pp$ collisions adequately.

The approach developed in this article also can be applied to study the production of other color-neutral particles or
heavy quark pair production in $pA$ collisions. One may expect that the CGC formalism and
 TMD factorization will
yield the same results for these processes in a certain kinematical region.
As a consequence, the  Weizs\"{a}cker-Williams and dipole type linearly  polarized gluon distributions
could be extracted by measuring azimuthal asymmetries
in these processes. We will address these issues in a forthcoming paper~\cite{AMSZ:prep}.

\noindent
{\bf Acknowledgments:}
This work has been supported by BMBF (OR 06RY9191).

%

% References

% =========

%


\begin{thebibliography}{99}



\bibitem{Boer:2011fh}

D.~Boer, M.~Diehl, R.~Milner, R.~Venugopalan, W.~Vogelsang, D.~Kaplan, H.~Montgomery and S.~Vigdor {\it et al.},
%``Gluons and the quark sea at high energies: Distributions, polarization, tomography,''
arXiv:1108.1713 [nucl-th].



%\cite{Mulders:2000sh}
\bibitem{Mulders:2000sh}
  P.~J.~Mulders and J.~Rodrigues,
 %``Transverse momentum dependence in gluon distribution and fragmentation
  %functions,''
  Phys.\ Rev.\  D {\bf 63}, 094021 (2001)
  [arXiv:hep-ph/0009343].
  %%CITATION = PHRVA,D63,094021;%%

  %\cite{Anselmino:2005sh}
\bibitem{Anselmino:2005sh}
  M.~Anselmino, M.~Boglione, U.~D'Alesio, E.~Leader, S.~Melis and F.~Murgia,
  %``The general partonic structure for hadronic spin asymmetries,''
  Phys.\ Rev.\ D {\bf 73}, 014020 (2006)
  [hep-ph/0509035].
  %%CITATION = HEP-PH/0509035;%%



  %\cite{Meissner:2007rx}
\bibitem{Meissner:2007rx}
  S.~Meissner, A.~Metz and K.~Goeke,
  %``Relations between generalized and transverse momentum dependent parton
  %distributions,''
  Phys.\ Rev.\  D {\bf 76}, 034002 (2007)
  [arXiv:hep-ph/0703176].
  %%CITATION = PHRVA,D76,034002;%%



%\cite{Boer:1997nt}
\bibitem{Boer:1997nt}
  D.~Boer and P.~J.~Mulders,
  %``Time-reversal odd distribution functions in leptoproduction,''
  Phys.\ Rev.\  D {\bf 57}, 5780 (1998)
  [arXiv:hep-ph/9711485].
  %%CITATION = PHRVA,D57,5780;%%



%\cite{Brodsky:2002cx}
\bibitem{Brodsky:2002cx}
  S.~J.~Brodsky, D.~S.~Hwang and I.~Schmidt,
  %``Final state interactions and single spin asymmetries in semiinclusive deep
  %inelastic scattering,''
  Phys.\ Lett.\  B {\bf 530}, 99 (2002)
  [arXiv:hep-ph/0201296].
  %%CITATION = PHLTA,B530,99;%%



%\cite{Collins:2002kn}
\bibitem{Collins:2002kn}
  J.~C.~Collins,
  %``Leading-twist Single-transverse-spin asymmetries: Drell-Yan and
  %Deep-Inelastic Scattering,''
  Phys.\ Lett.\  B {\bf 536}, 43 (2002)
  [arXiv:hep-ph/0204004].
  %%CITATION = PHLTA,B536,43;%%





%\cite{Metz:2011wb}
\bibitem{Metz:2011wb}
  A.~Metz, J.~Zhou,
  %``Distribution of linearly polarized gluons inside a large nucleus,''
  Phys.\ Rev.\  {\bf D84}, 051503 (2011).
  [arXiv:1105.1991 [hep-ph]].







%\cite{Boer:2009nc}
\bibitem{Boer:2009nc}
  D.~Boer, P.~J.~Mulders and C.~Pisano,
  %``Dijet imbalance in hadronic collisions,''
  Phys.\ Rev.\  D {\bf 80}, 094017 (2009)
  [arXiv:0909.4652 [hep-ph]].
  %%CITATION = PHRVA,D80,094017;%%



%\cite{Boer:2010zf}
\bibitem{Boer:2010zf}
  D.~Boer, S.~J.~Brodsky, P.~J.~Mulders and C.~Pisano,
  %``Direct Probes of Linearly Polarized Gluons inside Unpolarized Hadrons,''
  Phys.\ Rev.\ Lett.\  {\bf 106}, 132001 (2011)
  [arXiv:1011.4225 [hep-ph]].
  %%CITATION = PRLTA,106,132001;%%



%\cite{Qiu:2011ai}
\bibitem{Qiu:2011ai}
  J.~-W.~Qiu, M.~Schlegel and W.~Vogelsang,
  %``Probing Gluonic Spin-Orbit Correlations in Photon Pair Production,''
  Phys.\ Rev.\ Lett.\  {\bf 107}, 062001 (2011)
  [arXiv:1103.3861 [hep-ph]].
  %%CITATION = ARXIV:1103.3861;%%



  %\cite{Anselmino:2011ay}
\bibitem{Anselmino:2011ay}
  M.~Anselmino {\it et al.},
  %``Transverse Momentum Dependent Parton Distribution/Fragmentation Functions
  %at an Electron-Ion Collider,''
  Eur.\ Phys.\ J.\  A {\bf 47}, 35 (2011)
  [arXiv:1101.4199 [hep-ex]].
  %%CITATION = EPHJA,A47,35;%%





%\cite{Boer:2011kf}
\bibitem{Boer:2011kf}
  D.~Boer, W.~J.~den Dunnen, C.~Pisano, M.~Schlegel and W.~Vogelsang,
  %``Linearly Polarized Gluons and the Higgs Transverse Momentum Distribution,''
  Phys.\ Rev.\ Lett.\  {\bf 108}, 032002 (2012)
  [arXiv:1109.1444 [hep-ph]].
  %%CITATION = ARXIV:1109.1444;%%



  %\cite{Sun:2011iw}
\bibitem{Sun:2011iw}
  P.~Sun, B.~-W.~Xiao, F.~Yuan,
  %``Gluon Distribution Functions and Higgs Boson Production at Moderate Transverse Momentum,''
  Phys.\ Rev.\  {\bf D84}, 094005 (2011).
  [arXiv:1109.1354 [hep-ph]].



\bibitem{Manohar:2012jr}
  A.~V.~Manohar and W.~J.~Waalewijn,
  %``A QCD Analysis of Double Parton Scattering: Color Correlations, Interference Effects and Evolution,''
  arXiv:1202.3794 [hep-ph] and references therein.





\bibitem{GarciaEchevarria:2011rb}
  M.~Garcia-Echevarria, A.~Idilbi and I.~Scimemi,
  %``Factorization Theorem For Drell-Yan At Low q_T And Transverse Momentum Distributions On-The-Light-Cone,''
  arXiv:1111.4996 [hep-ph] and references therein.
  %%CITATION = ARXIV:1111.4996;%%









    %\cite{Catani:2010pd}
\bibitem{Catani:2010pd}
  S.~Catani, M.~Grazzini,
  %``QCD transverse-momentum resummation in gluon fusion processes,''
  Nucl.\ Phys.\  {\bf B845}, 297-323 (2011).
  [arXiv:1011.3918 [hep-ph]].



  %\cite{Mantry:2009qz}
\bibitem{Mantry:2009qz}
  S.~Mantry, F.~Petriello,
  %``Factorization and Resummation of Higgs Boson Differential Distributions in Soft-Collinear Effective Theory,''
  Phys.\ Rev.\  {\bf D81}, 093007 (2010).
  [arXiv:0911.4135 [hep-ph]]; and references therein.



    %\cite{Nadolsky:2007ba}
\bibitem{Nadolsky:2007ba}
  P.~M.~Nadolsky, C.~Balazs, E.~L.~Berger, C.~-P.~Yuan,
  %``Gluon-gluon contributions to the production of continuum diphoton pairs at hadron colliders,''
  Phys.\ Rev.\  {\bf D76}, 013008 (2007).
  [hep-ph/0702003 [HEP-PH]].


\bibitem{Collins:2011zzd}
  J.~Collins,
  %``Foundations of perturbative QCD,''
  (Cambridge monographs on particle physics, nuclear physics and cosmology. 32)




  %\cite{Hautmann:2002tu}
\bibitem{Hautmann:2002tu}
  F.~Hautmann,
  %``Heavy top limit and double logarithmic contributions to Higgs production at m(H)**2 / s much less than 1,''
  Phys.\ Lett.\  {\bf B535}, 159-162 (2002).
  [hep-ph/0203140].



 %\cite{Lipatov:2005at}
\bibitem{Lipatov:2005at}
  A.~V.~Lipatov, N.~P.~Zotov,
  %``Higgs boson production at hadron colliders in the k(T)-factorization approach,''
  Eur.\ Phys.\ J.\  {\bf C44}, 559-566 (2005).
  [hep-ph/0501172].

  %\cite{Pasechnik:2006du}
\bibitem{Pasechnik:2006du}
  R.~S.~Pasechnik, O.~V.~Teryaev and A.~Szczurek,
  %``Scalar Higgs boson production in a fusion of two off-shell gluons,''
  Eur.\ Phys.\ J.\ C {\bf 47}, 429 (2006)
  [hep-ph/0603258].
  %%CITATION = HEP-PH/0603258;%%



%\cite{Kovchegov:1996ty}
\bibitem{Kovchegov:1996ty}
  Y.~V.~Kovchegov,
  %``NonAbelian Weizsacker-Williams field and a two-dimensional effective color
  %charge density for a very large nucleus,''
  Phys.\ Rev.\  D {\bf 54}, 5463 (1996)
  [arXiv:hep-ph/9605446].
  %%CITATION = PHRVA,D54,5463;%%



%\cite{JalilianMarian:1996xn}
\bibitem{JalilianMarian:1996xn}
  J.~Jalilian-Marian, A.~Kovner, L.~D.~McLerran and H.~Weigert,
  %``The Intrinsic glue distribution at very small x,''
  Phys.\ Rev.\  D {\bf 55}, 5414 (1997)
  [arXiv:hep-ph/9606337].
  %%CITATION = PHRVA,D55,5414;%%



  %\cite{Dominguez:2010xd}
\bibitem{Dominguez:2010xd}
  F.~Dominguez, B.~W.~Xiao and F.~Yuan,
  %``$k_t$-factorization for Hard Processes in Nuclei,''
  Phys.\ Rev.\ Lett.\  {\bf 106}, 022301 (2011)
  [arXiv:1009.2141 [hep-ph]].
  %%CITATION = PRLTA,106,022301;%%



%\cite{Dominguez:2011wm}
\bibitem{Dominguez:2011wm}
  F.~Dominguez, C.~Marquet, B.~W.~Xiao and F.~Yuan,
  %``Universality of Unintegrated Gluon Distributions at small x,''
  Phys.\ Rev.\  D {\bf 83}, 105005 (2011)
  [arXiv:1101.0715 [hep-ph]].
  %%CITATION = ARXIV:1101.0715;%%



%\cite{McLerran:1993ni}
\bibitem{McLerran:1993ni}
  L.~D.~McLerran and R.~Venugopalan,
  %``Computing quark and gluon distribution functions for very large nuclei,''
  Phys.\ Rev.\  D {\bf 49}, 2233 (1994)
  [arXiv:hep-ph/9309289];
  %%CITATION = PHRVA,D49,2233;%%
%%\cite{McLerran:1993ka}
%\bibitem{McLerran:1993ka}
%  L.~D.~McLerran and R.~Venugopalan,
%  %``Gluon distribution functions for very large nuclei at small transverse
%  %momentum,''
  Phys.\ Rev.\  D {\bf 49}, 3352 (1994)
  [arXiv:hep-ph/9311205].
  %%CITATION = PHRVA,D49,3352;%%





%\cite{Mueller:2001fv}
\bibitem{Mueller:2001fv}
  A.~H.~Mueller,
  %``Parton saturation: An Overview,''
  arXiv:hep-ph/0111244.
  %%CITATION = HEP-PH/0111244;%%



%\cite{Iancu:2002xk}
\bibitem{Iancu:2002xk}
  E.~Iancu, A.~Leonidov and L.~McLerran,
  %``The Color glass condensate: An Introduction,''
  arXiv:hep-ph/0202270.
  %%CITATION = HEP-PH/0202270;%%





%\cite{Gelis:2010nm}
\bibitem{Gelis:2010nm}
  F.~Gelis, E.~Iancu, J.~Jalilian-Marian and R.~Venugopalan,
  %``The Color Glass Condensate,''
  Ann.\ Rev.\ Nucl.\ Part.\ Sci.\  {\bf 60}, 463 (2010)
  [arXiv:1002.0333 [hep-ph]].
  %%CITATION = ARNUA,60,463;%%







%\cite{Kuraev:1977fs}
\bibitem{Kuraev:1977fs}
  E.~A.~Kuraev, L.~N.~Lipatov, V.~S.~Fadin,
  %``The Pomeranchuk Singularity in Nonabelian Gauge Theories,''
  Sov.\ Phys.\ JETP {\bf 45}, 199-204 (1977).



%\cite{Gribov:1984tu}
\bibitem{Gribov:1984tu}
  L.~V.~Gribov, E.~M.~Levin, M.~G.~Ryskin,
  %``Semihard Processes in QCD,''
  Phys.\ Rept.\  {\bf 100}, 1-150 (1983).



    %\cite{Catani:1990eg}
\bibitem{Catani:1990eg}
  S.~Catani, M.~Ciafaloni, F.~Hautmann,
  %``High-energy factorization and small x heavy flavor production,''
  Nucl.\ Phys.\  {\bf B366}, 135-188 (1991).



%\cite{Collins:1991ty}
\bibitem{Collins:1991ty}
  J.~C.~Collins, R.~K.~Ellis,
  %``Heavy quark production in very high-energy hadron collisions,''
  Nucl.\ Phys.\  {\bf B360}, 3-30 (1991).



  %\cite{Dumitru:2010mv}
\bibitem{Dumitru:2010mv}
  A.~Dumitru, J.~Jalilian-Marian,
  %``Two-particle correlations in high energy collisions and the gluon four-point function,''
  Phys.\ Rev.\  {\bf D81}, 094015 (2010).
 [arXiv:1001.4820 [hep-ph]].
 %\cite{Dumitru:2010ak}
%\bibitem{Dumitru:2010ak}
  A.~Dumitru, J.~Jalilian-Marian,
  %``Forward dijets in high-energy collisions: Evolution of QCD n-point functions beyond the dipole approximation,''
  Phys.\ Rev.\  {\bf D82}, 074023 (2010)£¬
  [arXiv:1008.0480 [hep-ph]].



  %\cite{Dominguez:2011gc}
\bibitem{Dominguez:2011gc}
  F.~Dominguez, A.~H.~Mueller, S.~Munier, B.~-W.~Xiao,
  %``On the small-x evolution of the color quadrupole and the Weizs\'acker-Williams gluon distribution,''
  Phys.\ Lett.\  {\bf B705}, 106-111 (2011).
  [arXiv:1108.1752 [hep-ph]].





%\cite{Dominguez:2011br}
\bibitem{Dominguez:2011br}
  F.~Dominguez, J.~-W.~Qiu, B.~-W.~Xiao, F.~Yuan,
  %``On the linearly polarized gluon distributions in the color dipole model,''
  [arXiv:1109.6293 [hep-ph]].



  %\cite{Dumitru:2011vk}
\bibitem{Dumitru:2011vk}
  A.~Dumitru, J.~Jalilian-Marian, T.~Lappi, B.~Schenke, R.~Venugopalan,
  %``Renormalization group evolution of multi-gluon correlators in high energy QCD,''
  [arXiv:1108.4764 [hep-ph]].



  %\cite{Iancu:2011ns}
\bibitem{Iancu:2011ns}
  E.~Iancu, D.~N.~Triantafyllopoulos,
  %``Higher-point correlations from the JIMWLK evolution,''
    [arXiv:1109.0302 [hep-ph]].



    %\cite{Jmarian}
\bibitem{Jmarian}
  J.~Jalilian-Marian,
  %``On perturbative limits of quadrupole evolution in QCD at high energy,''
  [arXiv:1111.3936 [hep-ph]].

  %\cite{Iancu:2011nj}
\bibitem{Iancu:2011nj}
  E.~Iancu and D.~N.~Triantafyllopoulos,
  %``JIMWLK evolution in the Gaussian approximation,''
  JHEP {\bf 1204}, 025 (2012)
  [arXiv:1112.1104 [hep-ph]].
  %%CITATION = ARXIV:1112.1104;%%



  %\cite{JalilianMarian:1997gr}
\bibitem{JalilianMarian:1997gr}
  J.~Jalilian-Marian, A.~Kovner, A.~Leonidov, H.~Weigert,
  %``The Wilson renormalization group for low x physics: Towards the high density regime,''
  Phys.\ Rev.\  {\bf D59}, 014014 (1999).
  [hep-ph/9706377].
%\cite{JalilianMarian:1997jx}
%\bibitem{JalilianMarian:1997jx}
  J.~Jalilian-Marian, A.~Kovner, A.~Leonidov, H.~Weigert,
  %``The BFKL equation from the Wilson renormalization group,''
  Nucl.\ Phys.\  {\bf B504}, 415-431 (1997).
 [hep-ph/9701284].
%\cite{Iancu:2000hn}
%\bibitem{Iancu:2000hn}
  E.~Iancu, A.~Leonidov, L.~D.~McLerran,
  %``Nonlinear gluon evolution in the color glass condensate. 1.,''
  Nucl.\ Phys.\  {\bf A692}, 583-645 (2001).
  [hep-ph/0011241].
  %\cite{Ferreiro:2001qy}
%\bibitem{Ferreiro:2001qy}
  E.~Ferreiro, E.~Iancu, A.~Leonidov, L.~McLerran,
  %``Nonlinear gluon evolution in the color glass condensate. 2.,''
  Nucl.\ Phys.\  {\bf A703}, 489-538 (2002).
  [hep-ph/0109115].





  %\cite{Kopeliovich:1998nw}
\bibitem{Kopeliovich:1998nw}
  B.~Z.~Kopeliovich, A.~V.~Tarasov, A.~Sch\"{a}fer,
  %``Bremsstrahlung of a quark propagating through a nucleus,''
  Phys.\ Rev.\  {\bf C59}, 1609-1619 (1999).
  [hep-ph/9808378].



  %\cite{Kovchegov:1998bi}
\bibitem{Kovchegov:1998bi}
  Y.~V.~Kovchegov, A.~H.~Mueller,
  %``Gluon production in current nucleus and nucleon - nucleus collisions in a quasiclassical approximation,''
  Nucl.\ Phys.\  {\bf B529}, 451-479 (1998).
  [hep-ph/9802440].



  %\cite{Kharzeev:2003wz}
\bibitem{Kharzeev:2003wz}
  D.~Kharzeev, Y.~V.~Kovchegov, K.~Tuchin,
 %``Cronin effect and high p(T) suppression in pA collisions,''
  Phys.\ Rev.\  {\bf D68}, 094013 (2003).
  [hep-ph/0307037].



  %\cite{Dumitru:2001ux}
\bibitem{Dumitru:2001ux}
  A.~Dumitru, L.~D.~McLerran,
  %``How protons shatter colored glass,''
  Nucl.\ Phys.\  {\bf A700}, 492-508 (2002).
  [hep-ph/0105268].



  %\cite{Blaizot:2004wu}
\bibitem{Blaizot:2004wu}
  J.~P.~Blaizot, F.~Gelis, R.~Venugopalan,
  %``High-energy pA collisions in the color glass condensate approach. 1. Gluon production and the Cronin effect,''
  Nucl.\ Phys.\  {\bf A743}, 13-56 (2004).
  [hep-ph/0402256].



  %\cite{Gelis:2005pt}
\bibitem{Gelis:2005pt}
  F.~Gelis, Y.~Mehtar-Tani,
  %``Gluon propagation inside a high-energy nucleus,''
  Phys.\ Rev.\  {\bf D73}, 034019 (2006).
  [hep-ph/0512079].

%\cite{Avsar:2012hj}
\bibitem{Avsar:2012hj}
  E.~Avsar,
  %``TMD factorization and the gluon distribution in high energy QCD,''
  arXiv:1203.1916 [hep-ph].
  %%CITATION = ARXIV:1203.1916;%%


  %\cite{Balitsky:1995ub}
\bibitem{Balitsky:1995ub}
  I.~Balitsky,
  %``Operator expansion for high-energy scattering,''
  Nucl.\ Phys.\  {\bf B463}, 99-160 (1996).
  [hep-ph/9509348].



%\cite{McLerran:1998nk}
\bibitem{McLerran:1998nk}
  L.~D.~McLerran, R.~Venugopalan,
  %``Fock space distributions, structure functions, higher twists and small x,''
  Phys.\ Rev.\  {\bf D59}, 094002 (1999).
  [hep-ph/9809427].



   %\cite{Gelis:2001da}
\bibitem{Gelis:2001da}
  F.~Gelis, A.~Peshier,
  %``Probing colored glass via q anti-q photoproduction,''
  Nucl.\ Phys.\  {\bf A697}, 879-901 (2002).
  [hep-ph/0107142].



    %\cite{Blaizot:2004wv}
\bibitem{Blaizot:2004wv}
  J.~P.~Blaizot, F.~Gelis, R.~Venugopalan,
  %``High-energy pA collisions in the color glass condensate approach. 2. Quark production,''
  Nucl.\ Phys.\  {\bf A743}, 57-91 (2004).
  [hep-ph/0402257].





  %\cite{Fukushima:2007dy}
\bibitem{Fukushima:2007dy}
  K.~Fukushima, Y.~Hidaka,
  %``Light projectile scattering off the color glass condensate,''
  JHEP {\bf 0706}, 040 (2007).
  [arXiv:0704.2806 [hep-ph]].





  %\cite{Braun:2000wr}
\bibitem{Braun:2000wr}
  M.~Braun,
  %``Structure function of the nucleus in the perturbative QCD with N(c) ---> infinity (BFKL pomeron fan diagrams),''
  Eur.\ Phys.\ J.\  {\bf C16}, 337-347 (2000).
  [hep-ph/0001268].



  %\cite{Braun:2000bh}
\bibitem{Braun:2000bh}
  M.~A.~Braun,
  %``Inclusive jet production on the nucleus in the perturbative QCD with N(c) ---> infinity,''
  Phys.\ Lett.\  {\bf B483}, 105-114 (2000).
  [hep-ph/0003003].



  %\cite{OITS-166}
\bibitem{OITS-166}
  J.~C.~Collins and D.~E.~Soper,
  %``Parton Distribution and Decay Functions,''
  Nucl.\ Phys.\ B\ {\bf 194}, 445  (1982).
  %%CITATION = NUPHA,B194,445;%%



  %\cite{hep-ph/0503015}
\bibitem{hep-ph/0503015}
  X.~-d.~Ji, J.~-P.~Ma and F.~Yuan,
  %``Transverse-momentum-dependent gluon distributions and semi-inclusive processes at hadron colliders,''
  JHEP\ {\bf 0507}, 020  (2005)
  [hep-ph/0503015].
  %%CITATION = JHEPA,0507,020;%%







  %\cite{FillionGourdeau:2008ij}
\bibitem{FillionGourdeau:2008ij}
  F.~Fillion-Gourdeau, S.~Jeon,
  %``Wilson lines: Color charge densities correlators and the production of eta-prime in the CGC for pp and pA collisions,''
  Phys.\ Rev.\  {\bf C79}, 025204 (2009).
  [arXiv:0808.2154 [hep-ph]].







%\cite{AMSZ:prep}
\bibitem{AMSZ:prep}
  E. Akcakaya, A.~Metz, A. Sch\"{a}fer, J.~Zhou, in preparation.



\end{thebibliography}
\end{document}